\documentclass[useAMS,usenatbib]{mn2e}
\usepackage{txfonts}
\usepackage{graphicx}
\usepackage[mathscr]{eucal}
\newcommand{\be}{\begin{equation}}
\newcommand{\ee}{\end{equation}}
\newcommand{\bdm}{\begin{displaymath}}
\newcommand{\edm}{\end{displaymath}}
\newcommand{\aap}{{Astron. Astrophys.}}
\newcommand{\aaps}{{Astron. Astrophys. Suppl.}}
\newcommand{\apj}{{Astrophys. J.}}
\newcommand{\apjl}{{Astrophys. J. Lett.}}

\newcommand{\mnras}{{MNRAS}}

\newcommand{\aj}{{Astron.~ J.}}

\def\pdot {\dot P}

\def\msun{{\rm M}_{\odot}}

\def\mdot {\dot M_{\rm W}}

\def\ltsima{$\; \buildrel < \over \sim \;$}
\def\lsim{\lower.5ex\hbox{\ltsima}}
\def\gtsima{$\; \buildrel > \over \sim \;$}
\def\gsim{\lower.5ex\hbox{\gtsima}}
\def\hd {HD\,49798}

\def\bd     {BD\,+37$^{\circ}$442}
\def\bdd   {BD\,+37$^{\circ}$1977}
\def\bddd {BD\,+28$^{\circ}$4211}
\def\xmm {\emph{XMM-Newton}}

\title[X-ray wind emission in BD\,+37$^{\circ}$442]{The lack of X-ray pulsations in the extreme helium star  BD\,+37$^{\circ}$442 and its possible stellar wind X-ray emission}
\author[Mereghetti S. et al.]{Sandro Mereghetti$^{1}$\thanks{E-mail:
sandro@iasf-milano.inaf.it},   Nicola La Palombara$^{1}$, Andrea Tiengo$^{2,1,3}$, Paolo Esposito$^{4}$
  \newauthor  \\
    $^1$ INAF -- IASF Milano, Via E. Bassini 15, I-20133 Milano, Italy\\
   $^2$ Scuola Universitaria Superiore IUSS Pavia, Piazza della Vittoria 15, I-27100 Pavia, Italy\\
  $^3$ Istituto Nazionale di Fisica Nucleare, Sezione di Pavia, Via A. Bassi 6, I-27100 Pavia, Italy \\
  $^4$ Anton Pannekoek Institute for Astronomy, University of Amsterdam, Postbus 94249, NL-1090-GE Amsterdam, The Netherlands            }

\begin{document}
\date{Accepted 16 December 2016; Received 2 November  2016}
\pagerange{\pageref{firstpage}--\pageref{lastpage}}
\pubyear{}
\maketitle
\label{firstpage}
\begin{abstract}
We report the results of a new \xmm\ observation of the helium-rich hot subdwarf \bd\ carried out in February 2016. The possible periodicity at 19 s seen in a 2011 shorter observation is not confirmed, thus dismissing the evidence for a binary nature. This implies that the observed soft X-ray emission, with a luminosity of a few 10$^{31}$ erg s$^{-1}$, originates in \bd\ itself, rather than in an accreting neutron star companion.  The X-ray spectrum is well fit by thermal plasma emission with a temperature of 0.22 keV and non-solar element abundances. Besides the overabundance of He, C and N already known from optical/UV studies, the X-ray spectra indicate also  a significant excess of Ne. The soft X-ray spectrum and the ratio of X-ray to bolometric luminosity, L$_{\rm X}$/L$_{\rm BOL}\sim2\times10^{-7}$,  are similar to those observed in massive early type stars. This indicates that the mechanisms responsible for plasma shock-heating can work also in the  weak stellar winds (mass loss rates   $\mdot\leq10^{-8}$ $\msun$ yr$^{-1}$) of low-mass hot stars.

\end{abstract}


\section{Introduction}

The  luminous (L$\sim$2.5$\times10^4$ L$_{\odot}$) helium-rich  O-type star \bd\ is one of the few hot subdwarfs that have been detected in the X-ray band.  
X-rays from hot subdwarfs can have two  different origins: they can be produced by accretion onto a neutron star or white dwarf  companion or they can be emitted by shock-heated plasma in their stellar wind, as it occurs in early type stars of higher mass and luminosity. In both cases,  X-ray observations of hot subdwarfs are useful because they provide a way to study the relatively weak winds of these low-mass stars   \citep[see][for a review]{mer16r}.

X-rays from \bd\ were discovered   with an \xmm\ observation carried out in  2011  \citep{lap12}. 
The spectrum was fit  by the sum of a blackbody with temperature kT=45$\pm$10 eV plus a faint power-law  with photon index $\Gamma\sim$2,
giving a  luminosity  between $\sim10^{32}$  and $\sim10^{35}$ erg s$^{-1}$ (for d=2 kpc, \citet{bau95}). 
The large uncertainty on the luminosity, due to the poorly constrained spectral parameters,  left  open both  of the above possibilities for the origin of the X-ray emission in \bd .
The \xmm\ data also showed a  periodicity at 19.2 s  (with a statistical significance of 3.2$\sigma$) pointing to the presence of a compact companion.  
However,  no evidence for  a binary nature had been reported in the literature  \citep{fay73,kau80,dwo82} and
a  recent campaign of  high-resolution optical spectroscopy did not reveal  radial velocity variations \citep{heb14}. This means that either the orbital plane has a very small inclination and/or the orbital period is of the order of months, or that \bd\ is indeed a single star and the reported signal was spurious. 

Here we report the results of a longer follow-up \xmm\ observation of \bd\ that we obtained in order to clarify the origin of its X-ray emission.

 \begin{figure}
\includegraphics[width=7cm,angle=90]{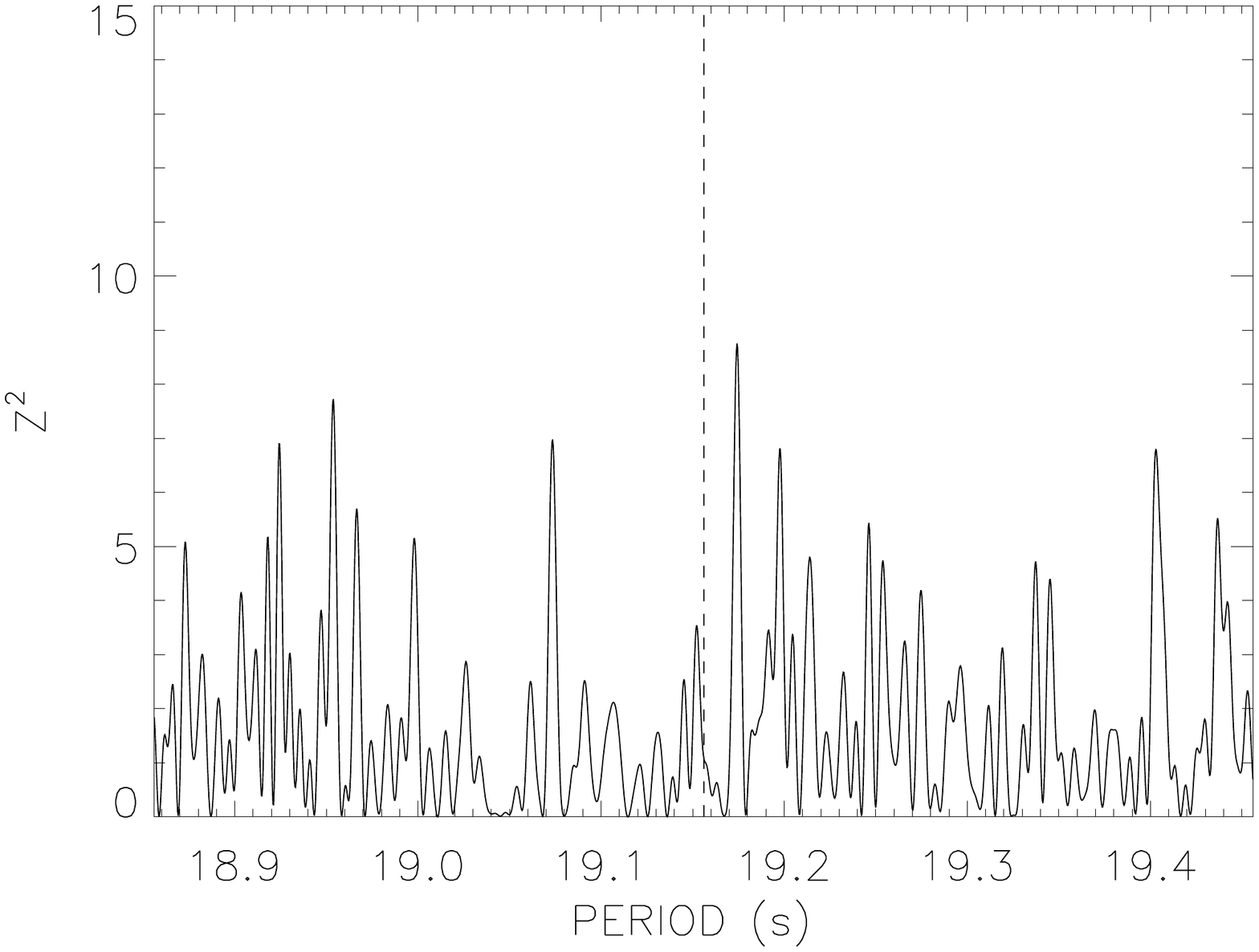}
\caption{Distribution of the Z$^2$ statistics vs. trial period for the 2016 observation of \bd . The dashed line indicates the period detected in the 2011 data. }
 \label{fig-z2}
 \end{figure}

 \begin{figure}
\begin{center}
\includegraphics[width=7cm,angle=-90]{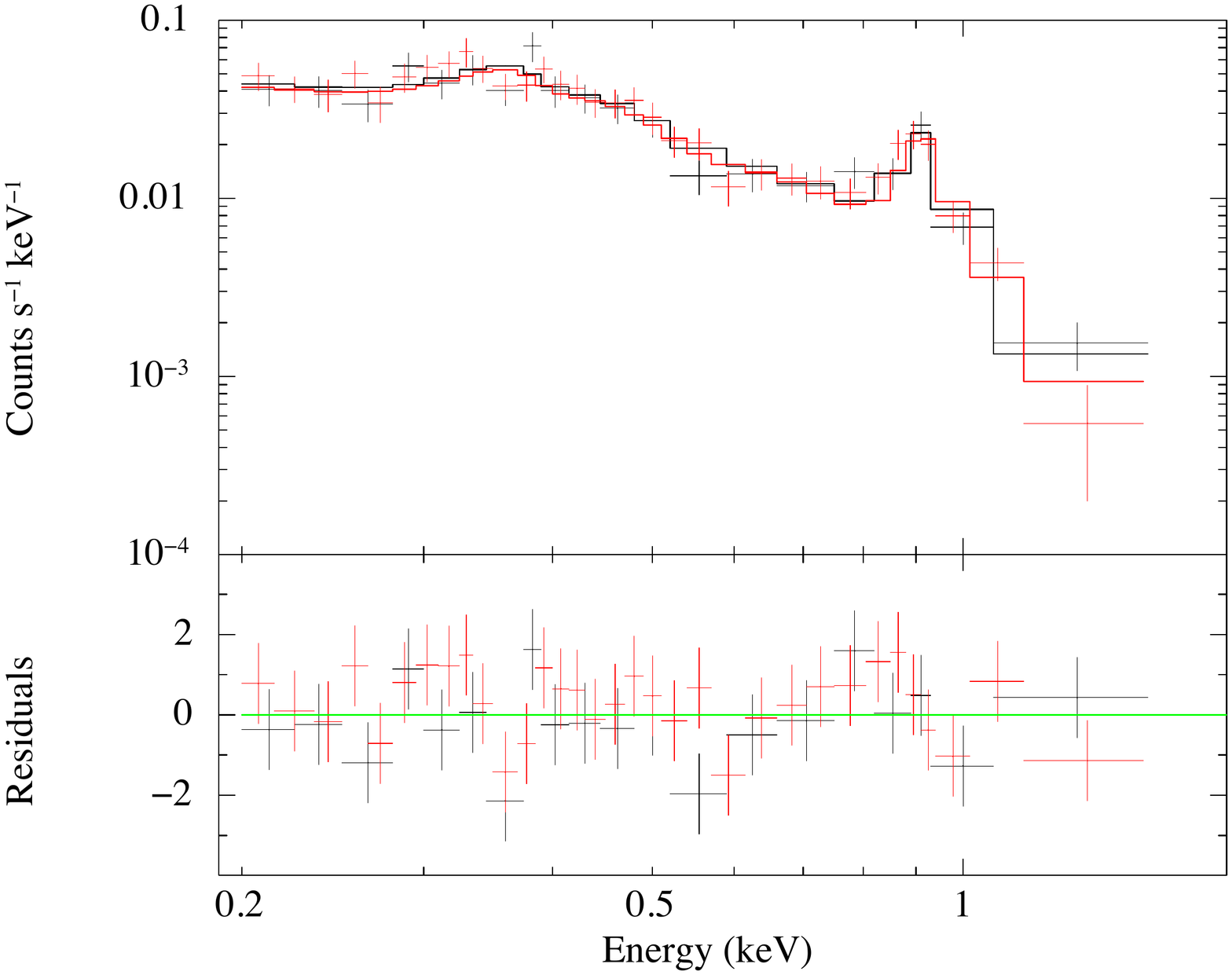}
\caption{Joint fit of the EPIC 2011 (black) and 2016 (red) spectra of  \bd\  with a thermal plasma model. The corresponding best-fit parameters are given in the   Model A column of Table 1. \textit{Top panel}: data and best fit model.  \textit{ Bottom panel}: residuals in units of $\sigma$. }
 \label{fig-sp}
 \end{center}
 \end{figure}

\section{Data analysis}
\label{analysis}
 
\bd\  was observed with the  \xmm\ satellite for 72 ks on 2016 February 1st. 
The three CCD cameras of the EPIC instrument were operated in imaging mode with a time resolution of 73 ms for the pn camera \citep{str01} and of 0.9    s for the two MOS cameras \citep{tur01}. The medium thickness filter was used for all cameras. The data were processed with SAS V15. We used only single- and double-pixel events for the pn (pattern $\leq$ 4) and single- and multiple-pixel events for the MOS (pattern $\leq$  12).

For the  timing analysis, we extracted the source counts from a circle of radius 15$''$ and converted their arrival times to the solar system barycenter. 
We used only the events in the  energy range from 0.15 to 2 keV, as it was done for the 2011 data by \citet{lap12}. This resulted in 1519 pn counts  and 485  MOS counts. We estimate that the background contributes  about 22\% and 17\% of these counts, respectively.  

In order to take into account  a possible spin-up or spin-down of the source during the   $\sim$4.5 years between the two \xmm\ observations with plausible values of  $|\pdot| \lsim 3\times10^{-10}$ s s$^{-1}$,  we considered possible periods in the range from 19.1 to 19.2 s  (the value measured in   2011   was P$_0$=19.156$\pm$0.001 s).    Using the sum of the pn and MOS counts, we found a maximum value of the Z$^2$ statistics \citep{buc83} of 8.74, for a period P=19.174 s  (see Fig.~\ref{fig-z2}).  However, the corresponding probability of chance occurrence, taking into account the number of sampled periods, is too large to claim a significant detection.
 
 Through Monte Carlo simulations we found that a sinusoidal modulation with pulsed fraction of 33\%, as found in the 2011 data,  has a probability of 99.9\% of being detected at a significance above 5$\sigma$ in an observation with the same duration and counting statistics as the 2016 one.  The corresponding probability for a pulsed fraction of 25\%  (the lower bound of the  2$\sigma$ uncertainty of the 2011 value) is 84\%.
Therefore, the lack of a detection in the new data strongly suggest    that either the pulsations in \bd\ disappeared (i.e. the pulsed fraction decreased making them undetectable) or   the   peak  at 19.2 s appearing in the  2011 periodogram was caused by a statistical fluctuation.
 
For the spectral analysis we used  circular extraction regions with radii of  20$''$ for the source and 50$''$ for the background. Time intervals of high background were excluded.  
We merged the  spectra from the three cameras into a single EPIC spectrum and produced  the appropriate response matrix using the task {\tt multixmmselect}.
Following exactly the same procedures, we extracted also the EPIC spectrum of \bd\ from the 2011 data, which we reprocessed using SAS v15. 

The effective exposure times of the   2016 and 2011 spectra are 48 ks and 28 ks,  respectively. 
By comparing the two spectra we found no evidence that   the flux or spectral shape changed between the two observations.  
Therefore, we performed the   spectral analysis by jointly fitting the 2011 and 2016 data in the 0.2-2 keV energy range using the XSPEC software. 

 Simple   models (power law, blackbody, thermal bremsstrahlung) modified by  interstellar absorption could not fit the data  (values of reduced $\chi^2_{\nu} >$2.5).  
Thermal plasma emission models, with abundances fixed at Solar values  \citep{and89} were also rejected, but they could give satisfactory fits if some of the abundances were let free to vary.
For simplicity, and considering that the quality of the data does not allow us to constrain a large number of spectral   parameters, we considered only  a single-temperature plasma  described by the {\tt apec} model.

We found strong evidence for an overabundance of  C and Ne,   the latter being required to fit the significant excess of emission visible in the spectra at $\sim$0.9 keV. 
The derived C and Ne  abundances were found to depend on that of He, which was however poorly constrained by the X-ray data. Therefore,  considering that \bd\ is an extreme He-rich star, we fixed the abundance of this element to a mass fraction of X$_{He}$=0.99. In this case we obtained a  good fit  with temperature kT=0.22 keV, absorption N$_H = 2.7\times10^{20}$   cm$^{-2}$, and solar abundances relative to hydrogen for all the other elements (see Fig.~\ref{fig-sp} and Model A   in Table~\ref{tab1} ). 
From optical studies we know, however, that also other elements are overabundant in this star. 
Indeed, an equally good fit was  obtained  with the  abundances of He, C, N, Si and Fe fixed at   the values of  \citet{jef10}, but also in this case an overabundance of Ne was required (Model B   in Table~\ref{tab1}).  
The two models result in slightly different values of the  unabsorbed  flux,  both of  a few $10^{-14}$ erg cm$^{-2}$ s$^{-1}$ (0.2-2 keV), which, for a distance of 2 kpc \citep{bau95}, correspond to   X-ray luminosites in the range $\sim(1.5-3)\times10^{31}$ erg s$^{-1}$. 

 We performed also spectral fits fixing the  interstellar absorption  at N$_H$=6.2$\times$10$^{20}$ cm$^{-2}$, based on the \bd\ reddening of E(B--V)=0.09 \citep{jef10} and the   relation derived by \citet{guv09}.
In the case of Model A, we obtained a good fit ($\chi^2_{\nu}$=1.03 ) with similar abundances and a slightly lower temperature kT=0.20$\pm$0.01 keV, while Model B resulted in a worse fit to the data ($\chi^2_{\nu}$=1.7).

 \begin{figure*}
\begin{center}
\includegraphics[width=14cm,angle=0]{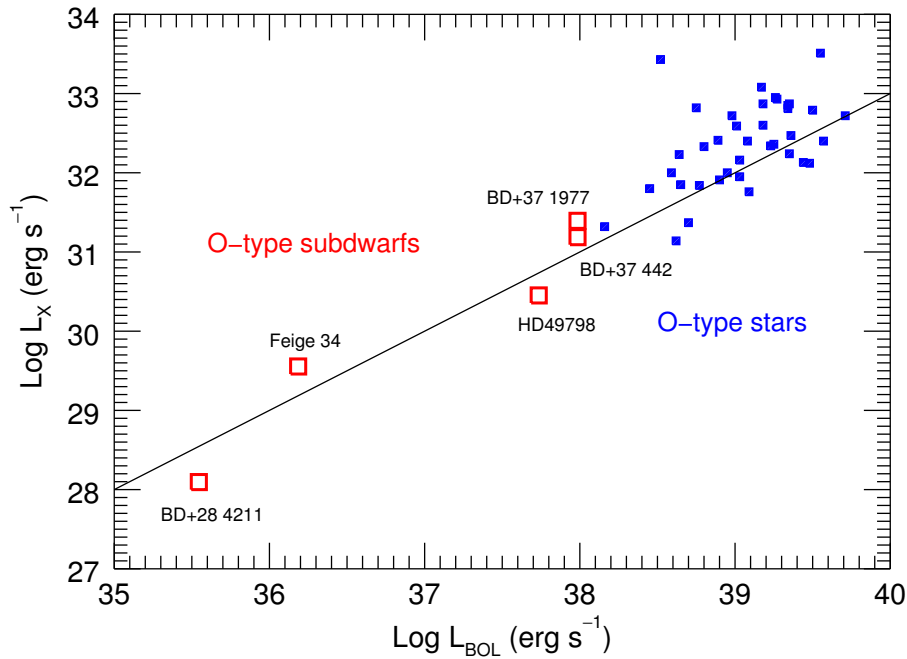}
\caption{X-ray versus bolometric luminosity for O-type subdwarfs (open red squares) and normal O-type stars (small blue squares). The line indicates the relation  $L_{\rm X}/L_{{\rm BOL}}=10^{-7} $.}
 \label{fig_lxlb}
 \end{center}
 \end{figure*}

\begin{table}
\caption{Results of the joint fits of the 2011 and 2014 EPIC spectra of \bd\ with two different assumptions for the element abundances.}
\label{tab1}
\begin{center}
\begin{tabular}{lcc} \hline
Parameter					& 	Model A				& 	Model B				\\ 
\hline
N$_H$ (10$^{20}$ cm$^{-2}$)   &  2.7$^{+1.4 }_{-0.9 }$ &  $<$7		 		\\
kT   (keV) 				&  	0.22$^{+0.01 }_{-0.02 }$	&  0.23$\pm$0.01			\\
X$_{\rm He}$        &    0.99 (fixed)                     &    0.96  (fixed) \\
X$_{\rm C}$         &       0.005                         &       0.025   (fixed) \\
X$_{\rm N}$         &                                   &      0.003 (fixed) \\
X$_{\rm Ne}$         &   0.0005     &  0.0025   \\
X$_{\rm Si}$         &       &  0.0008 (fixed)\\
X$_{\rm Fe}$         &       &  0.0006 (fixed)\\
Flux  (10$^{-14}$ erg cm$^{-2}$ s$^{-1}$) & 5.5$^{+1.1 }_{-0.6 }$  &3.4$^{+0.3 }_{-0.1 }$  	\\$\chi^2_{\nu}$/d.o.f.			& 0.94/48				& 0.90/49	 		\\ 
\hline
\end{tabular}
\end{center}

Notes: abundances are given in mass fraction; the flux refers to the 0.2-2 keV range and is corrected for the absorption; errors are at 1$\sigma$.
\end{table}

 \section{Discussion and Conclusions}
 \label{discussion}
 
The non-detection of a significant periodicity at 19.2 s in the new data (which have a higher counting statistics than the previous ones) casts doubts on the   presence of  a neutron star companion in \bd .    This result, coupled to the lack of radial velocity variations in the optical spectra \citep{heb14}, leads us to   conclude that \bd\ is most likely a single star.   Indeed, one of the scenarios proposed to explain the origin of extreme helium-rich stars is that they result from the merger of a binary composed of a He white dwarf and a more massive CO white dwarf. The lighter white dwarf is disrupted and a single  He-enriched star is formed \citep{jef11}. 

If \bd\ is a single star, the observed X-ray emission cannot be powered by mass accretion onto a neutron star companion, with the interesting consequence that  it has to originate in  \bd\ itself.
The X-ray emission from massive O-type stars is related to the presence of powerful radiation-driven winds. 
X-rays are produced in the winds of single stars by shock-heated plasma resulting from various instabilities \citep[see, e.g.,][and references therein]{osk16}. An empirical correlation has been found between the X-ray and bolometric luminosities of early type stars:   $L_{\rm X}\sim 10^{-7} ~L_{{\rm BOL}}$ \citep{pal81,naz09}. The X-ray luminosity we derived for  \bd\   
corresponds to a value of  $L_{\rm X}/L_{\rm BOL}\sim2\times10^{-7}$, fully consistent with the observed dispersion around the above average relation. 

It is thus natural to ascribe the X-rays observed in   \bd\ to the same, or similar, processes that are at work in the stellar winds of massive early type stars.      
There is in fact  evidence from  UV and optical spectroscopy  that \bd , despite its low luminosity compared to normal O-type stars,  has a stellar wind. The inferred mass-loss rate is $\mdot$=3$\times10^{-9}$ $\msun$ yr$^{-1}$ \citep{jef10}, in good agreement with the predictions of the most recent theoretical models of radiatively-driven winds in low luminosity hot stars \citep{krt16}.   

In Fig.~\ref{fig_lxlb} we plot the X-ray and bolometric luminosities of the five sdO stars that have been detected in the X-ray range. For comparison, the values for a sample of normal O stars seen in the ROSAT All Sky Survey \citep{ber96} are also shown in the figure.
\bdd\  \citep{lap15} is another single, He-rich  star very similar to \bd\ for what concerns its optical spectrum, composition and mass-loss properties \citep{jef10}. These two stars lie very close  in the $L_{\rm X}$-$L_{\rm BOL}$ plane. The same is true for the spectroscopic binary \hd , which is in a 1.55 day orbit with a neutron star or white dwarf companion \citep{mer09,mer13}.  For this star we plot in   Fig.~\ref{fig_lxlb} the X-ray luminosity  observed when the compact companion is eclipsed by the sdO star and likely due to wind emission; out of the eclipse the luminosity is a factor ten higher  and pulsations at 13.2 s are observed\footnote{  The presence of this X-ray periodicity in \hd\ is certain. The pulsations were discovered with high statistical significance in 1992 with the ROSAT satellite \citep{isr97} and subsequently detected in all the \xmm\ observations of this source carried out in  2002-2014 \citep{mer16}.}.  
The X-ray spectra of these three stars are similar, except for the presence of a  harder component in \hd\ which might be related to its binary nature. 
The two other sdO stars plotted in  Fig.~\ref{fig_lxlb} (Feige 34 and \bddd ; \citet{lap14}) are less luminous and their faint X-ray emission has not been studied in detail. It is however remarkable that they lie on the extrapolation of the  average L$_{\rm X}$-L$_{\rm BOL}$ relation observed at higher luminosity.  

 The temperature found in the spectral fits of \bd\ is in the lower range of the values seen  in the sample of   O-type stars observed with the  \xmm\ EPIC instrument, when a single-temperature plasma is sufficient to describe their low resolution X-ray spectra \citep{naz09}.  This might be related to the weaker wind of \bd\ compared to that of normal O-type stars.  It is also interesting to note that the weak-wind O-type dwarf $\mu$  Col (HD 38666), with a mass-loss rate   $\mdot$ = 10$^{-9.5\pm0.7}$   $\msun$ yr$^{-1}$ \citep{mar05} similar to that of \bd ,  has an  X-ray  emitting wind with  higher temperature (kT$\sim$0.4 keV, \citealt{hue12} ), despite its lower  wind velocity (v$_{\infty}$=1,200 km s$^{-1}$ wrt v$_{\infty}$=2,000 km s$^{-1}$, \citealt{jef10}). However, we caution that the temperatures derived from low-resolution  spectroscopy of faint X-ray sources,  such as the hot subdwarfs discussed here, might   be affected by the uncertainties in the element abundances. Instruments with large collecting area and  high spectral resolution are needed to address these issues in more detail and fully exploit the study of subdwarfs in the context of a  more general understanding of stellar wind in hot stars.


\section*{Acknowledgments}
This work,  supported  through financial contributions from the agreement ASI/INAF I/037/12/0 and from PRIN INAF 2014, is based on data from observations with {\it XMM-Newton}, an ESA science mission with instruments and contributions directly funded by ESA Member States and the USA (NASA). PE acknowledges funding in the framework of the NWO Vidi award A.2320.0076.

\bibliographystyle{aa}   

\bsp

\label{lastpage}

\end{document}